\documentclass[usegraphicx]{mn2e}

\usepackage{times}
\usepackage{amssymb}
\usepackage{amsmath}
\begin{document}

\include{defn}
\def\cm{{\rm\thinspace cm}}
\def\gm{{\rm\thinspace gm}}
\def\dyn{{\rm\thinspace dyn}}
\def\erg{{\rm\thinspace erg}}
\def\eV{{\rm\thinspace eV}}
\def\MeV{{\rm\thinspace MeV}}
\def\g{{\rm\thinspace g}}
\def\ga{{\rm\thinspace gauss}}
\def\K{{\rm\thinspace K}}
\def\keV{{\rm\thinspace keV}}
\def\km{{\rm\thinspace km}}
\def\kpc{{\rm\thinspace kpc}}
\def\Lsun{\hbox{$\rm\thinspace L_{\odot}$}}
\def\m{{\rm\thinspace m}}
\def\Mpc{{\rm\thinspace Mpc}}
\def\Msun{\hbox{$\rm\thinspace M_{\odot}$}}
\def\Zsun{\hbox{$\rm\thinspace Z_{\odot}$}}
\def\pc{{\rm\thinspace pc}}
\def\ph{{\rm\thinspace ph}}
\def\s{{\rm\thinspace s}}
\def\yr{{\rm\thinspace yr}}
\def\sr{{\rm\thinspace sr}}
\def\Hz{{\rm\thinspace Hz}}
\def\MHz{{\rm\thinspace MHz}}
\def\GHz{{\rm\thinspace GHz}}
\def\chisq{\hbox{$\chi^2$}}
\def\delchi{\hbox{$\Delta\chi$}}
\def\cmps{\hbox{$\cm\s^{-1}\,$}}
\def\cmpssq{\hbox{$\cm\s^{-2}\,$}}
\def\cmsq{\hbox{$\cm^2\,$}}
\def\cmcu{\hbox{$\cm^3\,$}}
\def\pcmcu{\hbox{$\cm^{-3}\,$}}
\def\pcmcuK{\hbox{$\cm^{-3}\K\,$}}
\def\dynpcmsq{\hbox{$\dyn\cm^{-2}\,$}}
\def\ergcmcups{\hbox{$\erg\cm^3\ps\,$}}
\def\ergpcmps{\hbox{$\erg\cm^{-3}\s^{-1}\,$}}
\def\ergpcmsqps{\hbox{$\erg\cm^{-2}\s^{-1}\,$}}
\def\ergpcmsqpspA{\hbox{$\erg\cm^{-2}\s^{-1}$\AA$^{-1}\,$}}
\def\ergpcmsqpspsr{\hbox{$\erg\cm^{-2}\s^{-1}\sr^{-1}\,$}}
\def\ergpcmcups{\hbox{$\erg\cm^{-3}\s^{-1}\,$}}
\def\ergpcmps{\hbox{$\erg\cm^{-1}\s^{-1}$}}
\def\ergps{\hbox{$\erg\s^{-1}\,$}}
\def\ergpspmp{\hbox{$\erg\s^{-1}\Mpc^{-3}\,$}}
\def\gpcm{\hbox{$\g\cm^{-3}\,$}}
\def\gpcmps{\hbox{$\g\cm^{-3}\s^{-1}\,$}}
\def\gps{\hbox{$\g\s^{-1}\,$}}
\def\Jy{{\rm Jy}}
\def\keVpcmsqpspsr{\hbox{$\keV\cm^{-2}\s^{-1}\sr^{-1}\,$}}
\def\kmps{\hbox{$\km\s^{-1}\,$}}
\def\kmpspmp{\hbox{$\km\s^{-1}\Mpc{-1}\,$}}
\def\Lsunppc{\hbox{$\Lsun\pc^{-3}\,$}}
\def\Msunpc{\hbox{$\Msun\pc^{-3}\,$}}
\def\Msunpkpc{\hbox{$\Msun\kpc^{-1}\,$}}
\def\Msunppc{\hbox{$\Msun\pc^{-3}\,$}}
\def\Msunppcpyr{\hbox{$\Msun\pc^{-3}\yr^{-1}\,$}}
\def\Msunpyr{\hbox{$\Msun\yr^{-1}\,$}}
\def\pcm{\hbox{$\cm^{-3}\,$}}
\def\pcmsq{\hbox{$\cm^{-2}\,$}}
\def\pcmK{\hbox{$\cm^{-3}\K$}}
\def\phpcmsqps{\hbox{$\ph\cm^{-2}\s^{-1}\,$}}
\def\pHz{\hbox{$\Hz^{-1}\,$}}
\def\pmpc{\hbox{$\Mpc^{-1}\,$}}
\def\pmpccu{\hbox{$\Mpc^{-3}\,$}}
\def\ps{\hbox{$\s^{-1}\,$}}
\def\psqcm{\hbox{$\cm^{-2}\,$}}
\def\psr{\hbox{$\sr^{-1}\,$}}
\def\kmpspMpc{\hbox{$\kmps\Mpc^{-1}$}}

\voffset -0.8in

\title{On the observed disc temperature of accreting intermediate mass
black holes}
\author[A.C. Fabian, R.R. Ross \& J.M. Miller]
{\parbox[]{6.in} {A.C. Fabian$^1$, R.R. Ross$^2$ and J.M. Miller$^3$\\
\footnotesize
1. Institute of Astronomy, Madingley Road, Cambridge CB3 0HA\\
2. Physics Department, College of the Holy Cross, Worcester, MA 01610,
USA\\
3. Harvard-Smithsonian Center for Astrophysics, 60 Garden St.,
Cambridge , MA 02138, USA\\
}}

\maketitle
\begin{abstract}
Some ultraluminous X-ray sources in nearby galaxies show soft
components resembling thermal disc emission. Calculations based on
blackbody emission then indicate that the accreting black holes at the
centres of these discs have masses of 100s to 1000s $\Msun$.
Establishing the existence of such intermediate mass black holes is of
considerable importance so the assumptions and approximations lying
behind blackbody spectral fits must be examined. We study here the
basic assumption that the thermal emission is well-characterised by a
multi-temperature blackbody spectrum. Since the opacity in the surface
layers of a disc decreases at high energies the emergent spectrum is
hardened. We compute the observed spectra from discs around a
non-spinning 1000$\Msun$ black hole and fit them over the XMM-Newton
pn band with multi-colour disc models. The typical {\em overall}
spectra hardening factor usually adopted for discs around stellar mass
black hole (assuming an inclination of 60~deg and including
relativistic effects for a non-spinning black hole) is found to be
appropriate.
\end{abstract}

\section{Introduction}

X-ray imaging of nearby galaxies often reveals the presence of
off-nucleus Ultra-Luminous X-ray sources (ULX for short) with X-ray
luminosities exceeding $10^{39}\ergps$ (for reviews see Fabbiano \&
White 2003; Miller \& Colbert 2003). Some have X-ray luminosities of
several times $10^{40}\ergps$. The variability of some ULX indicates
that they are single objects, most likely a form of accreting black
hole. The high luminosity, after a reasonable bolometric correction,
exceeds the Eddington limit for $20\Msun$, requiring a black hole of
mass above that expected from the collapse of normal massive stars.
The black hole mass inferred from the Eddington limit applied to the
more luminous objects is $\sim 500-1000\Msun$ and exceeds that of the
most massive stars known. 

The possibility that such Intermediate Mass Black Holes (IMBH) exists
is intriguing and raises questions about their origin and growth. Two
main scenarios have been proposed; the first considers them to be
remnants of the first stellar generation, Population III (Madau \&
Rees 2001), the second has them form recently by the merger of
stellar-mass black holes in a star cluster (Portegies Zwart et al
2004). The IMBH hypothesis is supported by the observation of giant
optical bubble nebulae around some ULX (Pakull \& Mironi 2002) and by
X-ray spectral evidence that some show evidence of cool accretion
discs, consistent with them being simply accreting black holes like
Cyg X-1 in its high state, scaled up in mass (Miller et al 2003, 2004.
Dewangan et al 2004; Cropper et al 2004). Strohmayer \& Mushotzky
(2003) have moreover found an iron line in the spectrum of the ULX in
M82.

Alternative hypotheses have been proposed. The first considers that
the emission is anisotropic. This can result from either collimation
at the source (King 2002) or relativistic beaming in a jet (Reynolds
et al 1997; ref). The other hypothesis proposes that super-Eddington
accretion is possible, i.e. that accretion can result in a
super-Eddington X-ray luminosity at infinity (e.g. Begelman 2002). 
Of course the ULX population need not be homogeneous and all
hypotheses may be represented by actual objects.

Here we remove one concern about the IMBH hypothesis. This is that the
correction factor applied to the thermal radiation from the accretion
disc to account for the high-energy tail departing from a perfect
blackbody spectrum taken from calculations for stellar mass objects
(Shimura \& Takahara 1994; Merloni, Ross \& Fabian 1999), might not
apply to IMBH. The high energy tail is the most detectable part of the
disc spectrum and is due to departures from a blackbody spectrum at
high photon energies by the innermost radii of the accretion disc (see
Laor \& Netzer 1989 for spectra expected from discs around
supermassive black holes). The observed spectrum is typically fitted
with a simple multi-colour disc (MCD) model appropriate for a
blackbody disc (Mitsuda et al 1984) and the black hole mass inferred
from the luminosity of that inner region, after the correction factor
has been applied. The correction factor is calculated here using the
same method as in the work of Merloni et al (1999). Model spectra have
been computed using constant-density disc models. These have then been
used to create fake X-ray spectra over the appropriate energy range
and then fitted with the MCD model. Our results show that current work
has estimated the black hole masses correctly provided that they are
non-spinning and have an inclination of about 60~deg. The required
mass rises as the black hole spin rises and drops slightly as the
inclination decreases.

\section{Method}

The calculation of the disc spectrum proceeds as described
by Merloni et al.\ (2000) with central mass $M=1000\,{\rm M}_\odot$.
At each radius, $r$, along the disc, the emergent spectrum is 
calculated assuming that the gas is fully ionized and that the 
gas density and the dynamic heating rate are uniform with depth. 
The density of the gas and the thickness of the disc at that 
radius are given by the formulae of Merloni et al.\ (2000).  
The disc is assumed to be radiation-pressure dominated for radii 
less than 36$R_{\rm S}$, 70$R_{\rm S}$ and 100$R_{\rm S}$ for 
accretion rates of 0.05, 0.10 and 0.20 times the Eddington limit, 
respectively. The dynamic heating rate is set so that the proper 
total emergent flux,
\begin{equation}
F=\frac{3GM\dot{M}}{8\pi r^3}\left(1-\sqrt{\frac{3R_{\rm S}}{r}}\right),
\end{equation}
is produced (Shakura \& Sunyaev 1973).

The radiative transfer and the vertical temperature structure at 
each radius are treated self-consistently, taking into account 
incoherent Compton scattering and free-free processes, as described 
by Ross \& Fabian (1996). (We do not use the coherent-scattering 
approximation in any of our calculations.) Typical emergent spectra
are computed for twenty annuli that cover the region
$3R_{\rm S}\leq r \leq 200R_{\rm S}$ of the disc. The effects of
relativistic blurring on the observed total spectrum are then
calculated using the method of Chen, Halpern \& Filippenko (1989)
with inclination $i=60^{\circ}$ assumed. An example is shown in Fig.~1.

\section{Results}

\begin{table}
\begin{tabular}{c c c c c c }
Accretion & $kT_{\rm in}$ (keV) & $kT_{\rm in}$ (keV) & $f_0$  & 
$kT_{\rm in}$ (keV) & $\beta$ \\
rate & \textsc{diskbb} & model & & \textsc{diskpn} &  \\
\\
0.05 & 0.209  & 0.100 & 1.40  & 0.199  & 1.28 \\
0.10 & 0.244  & 0.119 & 1.35  & 0.233  & 1.24 \\
0.20 & 0.284  & 0.142 & 1.30  & 0.271  & 1.19 \\
\end{tabular}
\caption{Results from fitting simulated data from the 3 model spectra
with the {\sc diskbb} and {\sc diskpn} MCD models. $f_0$ is the
overall hardening factor
required to convert the {\sc diskbb} normalization $K$ to $R_{\rm in}$
(eqn.~3).}
\end{table}

\begin{figure}
\includegraphics[angle=-90, width=0.95\columnwidth]{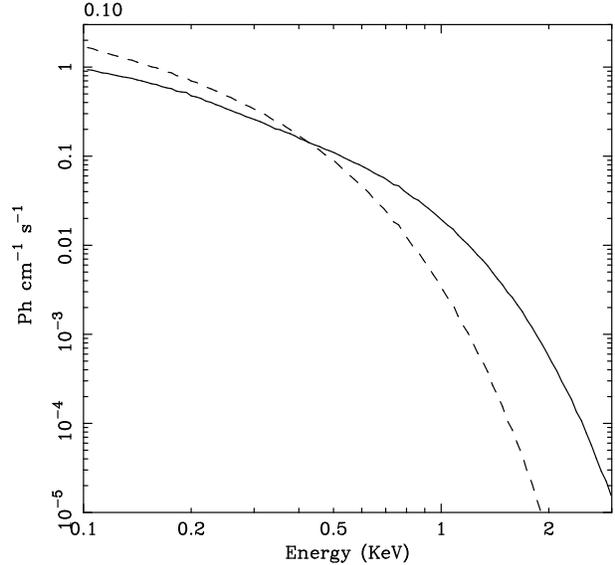}
\caption{Predicted spectrum, with first-order relativistic blurring, for a
disc accreting at a rate of 0.1 Eddington (solid line) compared with a
simple multi-temperature blackbody disc (dashed line) with the same
luminosity.}
\end{figure}

\begin{figure}
\includegraphics[angle=-90, width=0.95\columnwidth]{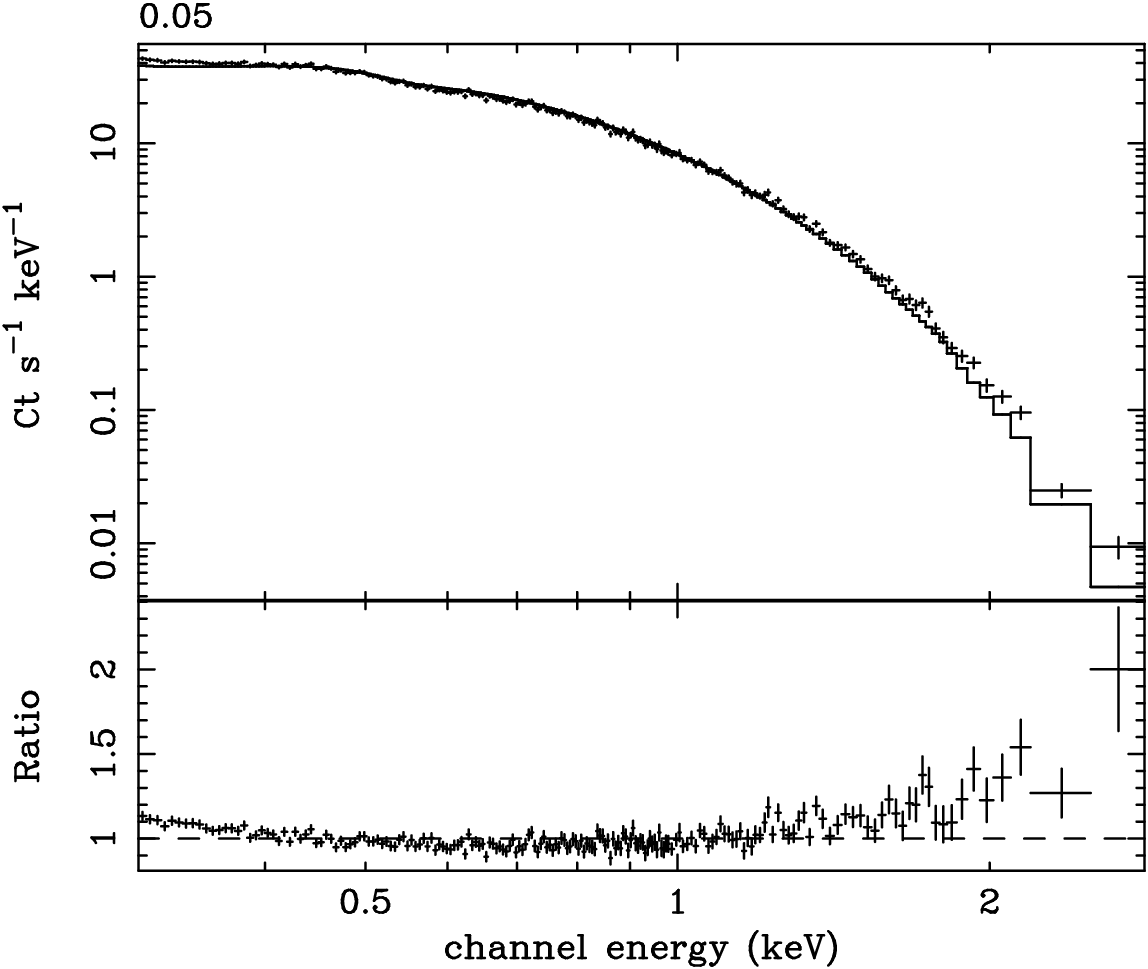}
\includegraphics[angle=-90, width=0.95\columnwidth]{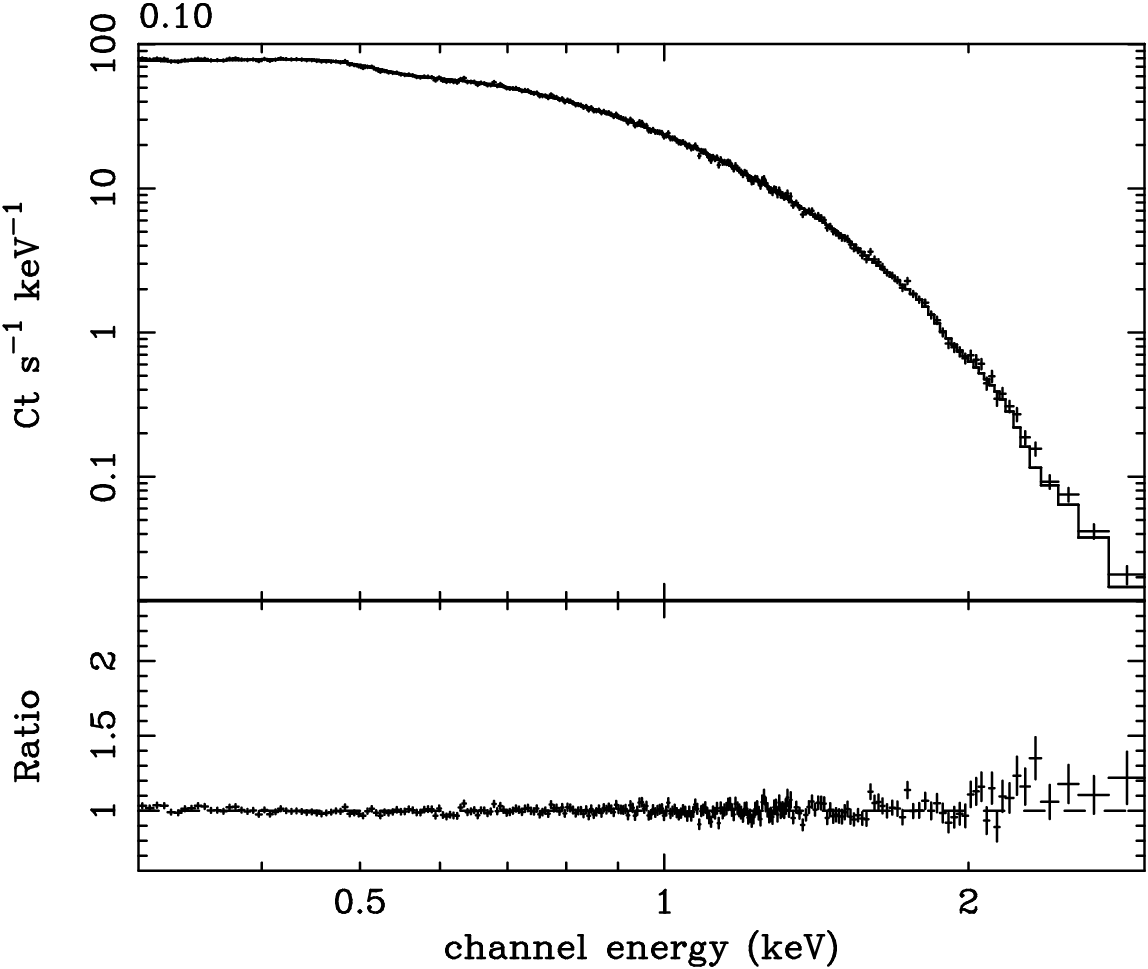}
\includegraphics[angle=-90, width=0.95\columnwidth]{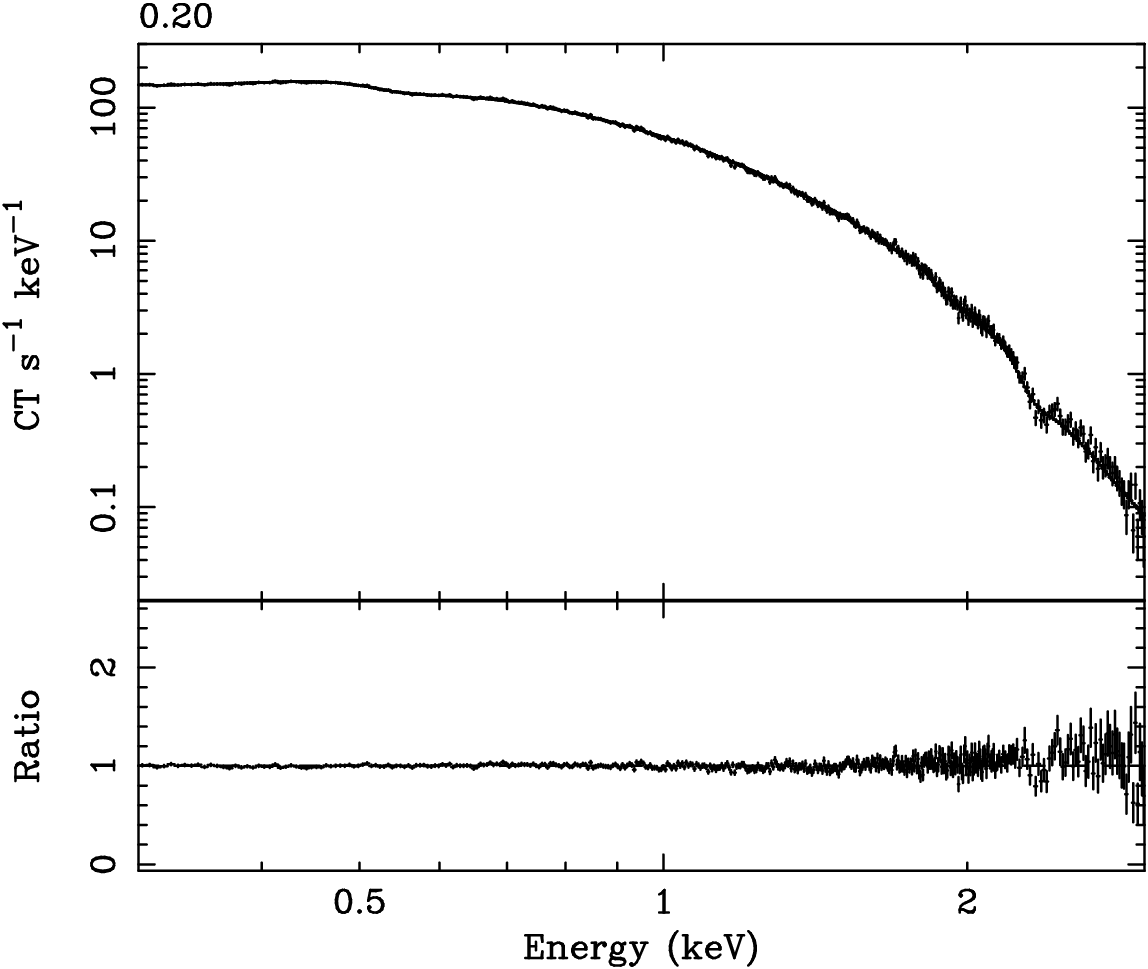}
\caption{Predicted spectra, simulated with the XMM-Newton pn response,
fitted with the simple multi-colour disc model \textsc{diskbb}. The fake
spectra are in the top panel and the ratio to the \textsc{diskbb} model in the
lower panel. Top figure: 0.05 Eddington, Middle: 0.1 Eddington and
Lower: 0.2 Eddington.}
\end{figure}

We now estimate spectral correction factors with which to convert the
results of spectral fitting of real data to physical parameters. The
correction factors will depend on the energy band over which they are
made and we therefore use that of the XMM-Newton EPIC pn CCD with
which several of the soft ULX spectra have been measured. In order to
do this we must fit our (blurred) model spectra with the multicolour
disc model commonly used.

The three model spectra were therefore used to fake XMM-Newton pn
spectral files which were then fitted over the 0.3--3~keV energy range
with the models \textsc{diskbb} (Fig.~2) and \textsc{diskpn}. The
results are shown in Table 1. The MCD model (\textsc{diskbb}; Mitsuda
et al 1984) assumes nothing about the inner boundary (there is no
zero-torque boundary condition at the innermost stable orbit of
$3R_{\rm S}$), whereas
\textsc{diskpn} (Gierli\'nski tet al 1999) assumes a zero-torque boundary
condition and uses a pseudo-Newtonian (pn) approximation.

The fitted inner disc temperature is about twice the equivalent
blackbody temperature of the inner disc for all 3 accretion rates
(Table 1) and both models, slightly exceeding the factor $f=1.7$
commonly used (Shimura \& Takahara 1994).

The black hole mass, $M_{\rm BH}$, is routinely obtained from the
normalization of the model, $K$, which essentially provides an
estimate of the emitting area. For the \textsc{diskbb} model the inner
radius of the disc (assumed to be $6GM_{\rm BH}/c^2$) is
\begin{equation}
R_{\rm in}=\eta g(i) D f^2 \sqrt{K/\cos i} \ \km,
\end{equation}
where $D$ is the distance to the source in units of $10\kpc$. $\eta$
and $g(i)$ account for (a) the fact that the disc emission peaks
outward of the innermost radius and (b) relativistic blurring and
beaming effects, respectively. Values of $\eta\sim0.5-0.7$, $g(i)\sim
0.7-0.9$ (Zhang, Cui \& Chen 1997) and $f=1.7$ (Shimura \& Takahara
1994) are often used. We have however included these effects in our
model so it is then appropriate to use $\eta g=1$ and $\cos i =0.5$,
yielding the values of $f_0$ shown in Table 1, where our $f_0^2=\eta g
f^2$. Our general result is 
\begin{equation}
R_{\rm in}={g(i)\over {g(60^{\circ})}}D f_0^2 \sqrt{K\over{\cos i}} \ \km,
\end{equation}
Miller et al (2003), for
example, assumed $\eta=0.63$ and $f=1.7$ which combine to give an
overall correction factor very close to our
calculated correction factor $f_0=1.35$ for an accretion rate of 0.10
times the Eddington limit. Those observational
results are therefore robust within the assumptions of our model.

The \textsc{diskpn} model (Gierli\'nski et al 1999) has a normalization
defined so that
\begin{equation}
M_{\rm BH}=D \beta^2 \sqrt{K/\cos i} ,
\end{equation}
where the distance $D$ is in kpc and the spectral hardening factor is
now denoted $\beta$. Values of $\beta$ deduced from our model fit are
listed in Table 1.

\section{Discussion}
We have shown that simple MCD models which assume blackbody spectra
are a good fit in the XMM-Newton pn CCD band for the disc emission
from accreting IMBH. The fitted inner-disc temperature is about a
factor of two larger than the actual inner temperature. An {\em
overall} spectral hardening factor of $f_0=\sqrt{\eta g}f\approx1.35$
is appropriate for estimating the black hole mass from the
normalization of the spectral fit. The masses reported from recently
modelling of the observed soft X-ray spectra of some nearby ULX are
robust provided that the observed inclination is about 60~deg. The
required mass decreases slightly if the inclination is smaller than
this.

The hard tail to the disk spectrum is mostly the result of the surface
layers of the inner disc becoming optically thin at higher energies.
Photons from deeper, hotter, subsurface regions escape leading to more
high energy emission than expected from a blackbody at the surface
temperature. Comptonization tends to slightly reduce this tail due to
scattering in the intervening regions between the photon source and
the surface (Ross et al 1992). Remarkably, the spectral correction
factor for XMM-Newton CCD spectra over the 0.3--3~keV range found here
for 1000$\Msun$ black holes is similar to that found for RXTE spectra
in the 2--20~keV band for 10$\Msun$ black holes. Further effects such
as more dissipation in the outer layers will increase the hard tail
making the correction factor larger.

Our results confirm that the detection of cool accretion discs in ULX
is strong evidence that they contain IMBH.

\section{Acknowledgements}

ACF thanks the Royal Society for support and JMM acknowledges support
from the NSF through its Astronomy and Astrophysics Fellowship
program.

\end{document}